\documentclass[english,format=sigconf]{acmart}
\usepackage{babel}
\usepackage{enumitem}
\AtBeginDocument{%
  }

\setcopyright{none}
\acmDOI{}
\acmISBN{}

\acmConference[LOCO2026]{2nd International Workshop on Low Carbon Computing}{Sept 10--11, 2026}{Lancaster \& Online}

\usepackage[T1]{fontenc}
\usepackage{graphicx}

\begin{document}

\title{Estimating the growth in emissions from AI data centres}

\author{Wim Vanderbauwhede}
\email{wim.vanderbauwhede@gla.ac.uk}

\affiliation{%
  \institution{University of Glasgow}
  \city{Glasgow}
  \state{}
  \country{UK}
}

\begin{abstract}
The advent of agentic AI is driving an unprecedented growth in data centre expansion. Using state-of-the-art life cycle assessment models for operational and embodied emissions from AI servers, we estimate the ensuing growth in the overall CO$_2$ emissions of AI data centres. 
The main contribution of this work is a rigorous quantification of emissions arising from projected AI data centre expansion using scenarios by the IEA and McKinsey \& Company. We show that the scenarios promoted by the AI industry would result in a dramatic rise in overall emissions and that the embodied carbon component is considerable.

\end{abstract}

\maketitle

\section{Introduction}

Climate change is an environmental problem. But our environment is
what allows our society to thrive. The damage from climate change
is therefore societal and economical as well as ecological, e.g. \cite{ukactuaries2025climate} predicts a 50\% drop in GDP at 2.5\textdegree C. The only way
to minimise this damage is to reduce global CO\textsubscript{2} emissions.
Even keeping them at the current level will cause catastrophic warming.
All this is explained in detail in the 2024 UNEP Emissions Gap Report \cite{ref16}.

The current push for generative AI in general and ``agentic'' AI in particular is deeply problematic in many ways.
In this paper we focus on CO$_{2}$ emissions arising not only from
the use of AI technology (operational emissions) but also the emissions
resulting from the building of the data centres and the manufacturing
of the servers, the so-called embodied or embedded emissions. 

The main technical contribution of this paper is the incorporation into the LCA
model of the trends of the contributions to the embodied emissions
of the servers. In this way our work improves on the state of the
art \cite{wadenstein2025life} where the values of embodied emissions
of the servers are static. 

The main contribution of the paper to the ``AI'' debate is a rigorous scenario-based quantification of emissions arising from AI data centre expansion.

\section{Emissions for projected growth }

Let's first consider the very real environmental damage caused not
so much by the AI technology itself as by the hype surrounding it. By
``hype'' we mean the excessive promotion of a technology. The purpose of hype in venture capital
based technology adoption is well-studied \cite{RADY2025106499,DEDEHAYIR201628}.
We want to consider here the \emph{effect} of the current AI hype.
\begin{itemize}
\item The hype creates an expectation of huge growth in demand for AI. ``Agentic'' AI is the latest manifestation of this hype, and it requires thousands times more resources than chatbot-style generative AI. 
This growth in data centre capacity is seen as a good thing, and of course the AI companies do not mention the concomitant emissions. Data
centre companies have to start building capacity before the demand
is realised. As a result, data centre capacity is being built up right now at an unprecedented scale.
\item This requires electricity generators to provision capacity for those
future data centres. It also requires
provisioning of semiconductor fab capacity.
\item Without strong growth in demand, electricity generators would phase
out fossil fuel generation because generating electricity from renewable
sources is more cost-effective. Because of the AI hype, they are no
longer phasing out fossil fuel generation as they want to maximise
generation capacity to maximise future profits. New fossil fuel powered
electricity plants are being developed as a result \cite{ref1} and
existing ones are kept open for longer \cite{ref2}.
\item As electricity generators of course want to optimise current profits
as well, they want to sell all the electricity they can generate,
rather than let plants idle. For the same reason, if semiconductor
fab capacity is increased, more semiconductor wafers must be produced
and therefore sold.
\item And so global emissions from electricity generation are not decreasing
at all, and are even expected to rise in the near future. This comes at
a time when we need to reduce global emissions urgently and drastically.
Embodied emissions from semiconductor manufacturing are increasing
with the increased volume. The extent of this increase depends on
the trend in embodied emissions.
\end{itemize}

Because of this very real effect of the AI hype on global emissions, it is important
to quantify the emissions arising from the \emph{projected} growth
in AI use, both embodied and operational.

\section{Breakdown of data centre emissions}

The greenhouse gas emissions from a data centre can be broken down
into a few main components, based on when and where the emissions
are incurred:
\begin{enumerate}
\item Emissions incurred while building the data centre infrastructure (mainly
the building itself and the cooling system). This is a minority share yet constitutes an important contribution to the overall embodied carbon.
\item Emissions incurred while manufacturing the servers used in the data
centre. The part of server manufacturing that produces by far the
most emissions is the chip production. Our model includes the carbon costs
of the packaging, enclosures etc. as well.
\item Emissions incurred while producing the electricity to run the servers.
This is called the carbon intensity of electricity generation and
depends on the location of the data centre. As AI data centres are
globally distributed, we will consider the global average emissions
of electricity production.
\end{enumerate}
We call (1) and (2) the \textquotedblleft embodied (carbon) emissions\textquotedblright{}
and (3) the \textquotedblleft operational (carbon) emissions\textquotedblright .

The global carbon intensity of electricity generation is decreasing at a rate of about 2.5\% per year. Unfortunately, this has not resulted in a decrease in emissions, which are still \emph{rising} by 2.3\% per year: what we see currently is that renewables are installed in addition to fossil fuel generation, rather than replacing them. Power generation CO$_2$ emissions are plateauing rather than reducing steeply \cite{ref2b}. 

\section{An estimate of the growth in emissions}

We have created a model for the evolution of data centre emissions over time, which takes into account both the embodied carbon and the emissions from use. This model does also take into account the embodied carbon
emissions from creating the actual data centres infrastructure but
not the supporting infrastructure (electricity supplies, networking,
roads, water supplies).

\subsection{Data centre LCA model}

Our LCA model for data centres was published in \cite{wadenstein2025life}. We
 refer to that work for full details but present here a summary
of the model construction.

The total carbon footprint for a given capacity over time \emph{t}
is the sum of the embodied carbon of the hardware (including manufacturing,
transport and end-of-life disposal) and the operational emissions (Eq. \ref{eq1}). In all following equations, $t \in \mathbb{N}$.
\begin{eqnarray}
\mathtt{total\_emissions}(t) ~ = & ~ \\
& \mathtt{operational\_emissions}(t)\nonumber \\
& +\mathtt{embodied\_emissions}(t)\nonumber 
\label{eq1}
\end{eqnarray}

\subsubsection{Site parameters}

The emissions depend on the site in various ways. The factors taken into account are:

\begin{itemize}
    \item yearly site expansion factor, $\alpha$
    \item yearly average site load factor, $\lambda=0.8$    
    \item site power usage effectiveness, $\textit{PUE}$
    \item site node lifetime, $\Delta t$ (years) 
    \item \#nodes in data centre, $n_{nodes}$
    \item node embodied carbon, EC$_{node}$ (kgCO$_2$e) 
    \item node power consumption, $P_{node}$ (W)
    \item node idle power consumption factor, $\gamma=0.3$
    \item location electricity~carbon~intensity $CI$ (kgCO$_2$e/kWh) 
\end{itemize}

The node energy consumption per year, $E_{node}$ (kWh/year), takes into account the load and the idle power consumption:
\begin{equation}	\label{eq1b}
	E_{node} = (\lambda+(1-\lambda )\cdot\gamma)\cdot P_{node}\cdot(24\cdot 365/1000)
\end{equation}

The node embodied carbon is discussed in Section \ref{sec:embodied-emissions}. 

\subsubsection{Data centre expansion and server replacement}

In practice, the data centre will periodically renew its hardware
and also expand its capacity. This has an effect on the embodied
and operational emissions as replacing hardware increases
the embodied carbon; but though the newer generation hardware
is more energy efficient, it does no lead to reductions in operational emissions because the data centre is power-capped. Our model includes a correction for the energy efficiency of computation (Koomey's law \cite{koomey2011web}), but in this 
paper we therefore omit it.

We model the expansion as

\begin{equation}\label{eq2}
\mathtt{expansion}(t,\alpha) = (1+\alpha)^{t - 1} 
\end{equation}
~\\
The replacement is conceptually modelled using a generic step function with $\Delta t$ the replacement period:

\begin{equation}\label{eq3}
\mathtt{step}(t,\Delta t) =  \Delta t \cdot \left\lfloor \frac{t-1}{\Delta t}\right\rfloor + 1 
\end{equation}

(This function returns $1+(n-1) \cdot \Delta t$ for all values in the $n^{th}$ period.)

\subsubsection{Operational emissions}\label{sec:operational-emissions}

The operational emissions depend in first order on the electricity
consumption of the hardware, the power usage effectiveness (PUE) of the
data centre and the electricity carbon intensity (amount of CO$_2$ emitted
per unit of electricity produced) of the electricity generation.

Furthermore, carbon intensity should improve over time and depends on the
geographical area where the centre is located. 

The function expressing this is (Eq. \ref{eq5}):
\begin{eqnarray}\label{eq5}
\mathtt{carbon\_intensity\_correction}(t,\kappa)= \\ 
(1-\kappa ) ^{t-1}\nonumber 
\end{eqnarray}

So that
\begin{eqnarray}\label{eq6}
\mathtt{carbon\_intensity}(t,\kappa) & = &  ~ \\ 
CI \cdot \mathtt{carbon\_intensity\_correction}(t,\kappa)\nonumber & ~ & ~
\end{eqnarray}

With these assumptions, the expression for the data centre operational emissions per year becomes (Eq. \ref{eq7}):
\begin{eqnarray}\label{eq7}
\mathtt{operational\_emissions\_year}(t) = ~ ~\\
    n_{nodes} \cdot E\_{node} \nonumber \\
 \cdot(\mathtt{carbon\_intensity}(t,\kappa)) \cdot \textit{PUE}   \nonumber \\
 \cdot\mathtt{compute\_efficiency\_correction}(step(t,\Delta t),\beta) \nonumber\\
 \cdot\mathtt{expansion}(t,\alpha) \nonumber
\end{eqnarray}

Cumulative emissions from electricity consumption are obtained by summing the yearly emissions (Eq. \ref{eq7b}):
\begin{eqnarray}\label{eq7b}
\mathtt{operational\_emissions}(t) ~=~ \\
\sum_{t_c=1}^{t} \mathtt{operational\_emissions\_year}(t_c) \nonumber
\end{eqnarray}

\subsubsection{Embodied emissions}\label{sec:embodied-emissions}

For the embodied emissions of the site, we accumulate the embodied carbon of the new replacement cycle to the total embodied carbon at the time. As we replace all servers after a lifetime $\Delta t$, and we have a yearly cluster expansion, we get a combination between a stepwise replacement and the gradual replacement due to yearly expansion. The resulting expression for the embodied emissions over time is a recursive function. The model is constructed as follows:

\begin{itemize}
\item $n_{inst}$ is the number of nodes installed in a given year. 
This is the initial number multiplied with the expansion factor year by year, in other words the expansion is exponential.
\begin{equation}
\begin{aligned}
n_{inst}(t)~=~
	n_{init} \cdot 
	(1+\alpha)^t
\end{aligned} \label{eq10a}
 \end{equation}
\item $n_{new}$ is the new servers installed in a given year. 
This is made up of the additional nodes due to the yearly cluster expansion plus the retired nodes that have to be replaced.
\begin{equation}
\begin{aligned}
n_{new}(t,\Delta t)~=~\begin{cases}
n_{init} & ,~t=0\\
n_{inst}(t)+n_{ret}(t,\Delta t) & ,~t>0
\end{cases}
\end{aligned} \label{eq10b}
 \end{equation}
\item $n_{ret}$ is the number of retired server nodes in a given year. This is equal to the number of server nodes installed $\Delta t$ years before.
 \begin{equation}
\begin{aligned}
n_{ret}(t,\Delta t) & =\begin{cases}
0 & ,t<\Delta t\\
n_{new}(t-\Delta t,\Delta t) & ,t\geq\Delta t
\end{cases}
\end{aligned} \label{eq10c}
 \end{equation}
\end{itemize}
The equation for the embodied emissions is shown in Eq. \ref{eq10}. The embodied carbon corresponds to the new nodes, so we have
\setlength{\abovedisplayskip}{3pt}
\setlength{\belowdisplayskip}{3pt}
\begin{eqnarray}\label{eq10}
\mathtt{embodied\_emissions\_year}( t, \Delta t , EC_{node})~=~\\
EC_{node}  \cdot\, n_{new}(t,\alpha,\Delta t) \nonumber\\ 
 \cdot\, \mathtt{embodied\_carbon\_correction}(t)\nonumber
\end{eqnarray} 

The trend in embodied carbon with every new hardware generation, $\mathtt{embodied\_carbon\_correction}(t)$, is the main contribution of this paper, and discussed in detail in Section \ref{sec:Embodied-carbon-trends}.

The final cumulative embodied carbon model (Eq. \ref{eqCumEmbEm}) includes two more factors to account for the data centre infrastructure, $EC_{node,if}$ which is the per-node portion of the embodied carbon of the construction of the data centre, and $\Delta EC_{node,if}$ which is the per-node portion of the yearly additional embodied carbon arising from  infrastructure maintenance. The values are based on \cite{schneider2023}.
\setlength{\abovedisplayskip}{3pt}
\setlength{\belowdisplayskip}{3pt}
\begin{eqnarray}\label{eqCumEmbEm}
\mathtt{embodied\_emissions}(t) ~=~ \\
\sum_{t_c=1}^{t} ( \mathtt{embodied\_emissions\_year}(t_c) \nonumber\\
+ ~\Delta EC_{node,if} \cdot t_c ~) + ~EC_{node,if} \nonumber
\end{eqnarray}
\subsection{Embodied carbon model}\label{subsec:Embodied-carbon-model}

Our model for the embodied carbon of servers and as data
centre infrastructure is also part of our work \cite{wadenstein2025life}. We present here a summary of the model construction, so that it is clear where the contribution of this paper fits in.

The factor $EC_{node}$ for the contribution of the embodied carbon of the hardware in Eq. \ref{eq10} needs to be calculated separately. 
The data available for embedded carbon in a compute node that comes
from manufacturing, transport, and end-of-life disposal is available for a few
select server models from hardware vendor reports. None of these servers are
typical compute nodes for scientific computing.

To compute the embodied carbon of server manufacturing, we have re-implemented the model by Boavizta \cite{lorenzini2021digital}. This model is very comprehensive but as it was published in 2021 and has not been updated since, we looked for more up-to-date estimates for various parameters. Beside estimates for the chips, the model also includes contributions from packaging, power supply, server enclosure and rack enclosure. 

In particular for the various chips used in the server (CPU, RAM, SSD) we use the ACT methodology (Architectural Carbon modelling Tool) \cite{10.1145/3470496.3527408}. This tool uses the electricity consumption of the manufacturing process, the embodied carbon for the materials, and the greenhouse gas potential for the various gases used in production, and combines this with the die size to obtain an estimate for the embodied carbon of the chip. We updated some of the parameters (those that were included without reference in the ACT paper) using data from \cite{10.1145/3630614.3630616}, \cite{9372004} and \cite{AnandTech2019}. We also extended the model to include non-integrated GPUs.

The model is based on the die size of the various chips in a server (CPU, GPU, RAM, SSD). The embodied emissions are calculated using process-specific data (embodied emissions for the materials and gases used, amount of electricity used, process yield) and the CI of the electricity used by the fab. For example, for the CPU we have Eq. \ref{eqEmbCarbonCPU}. The first line are the emissions from production of the die, which consist of emissions from the electricity used $en_p*ci_{fab}$, embodied emissions from the materials used in production $ec_m$ and embodied as well as incurred emissions from the gases used in production $ec_g$ (many of which have a much higher GWP than CO$_2$), corrected by the wafer yield. The second line takes into account the number of cores $n_{\textit{cores}}$ and units $n_{\textit{units}}$ in the CPU packages, as well as the embodied carbon of the base $ec_{\textit{base}}$ and of the package $( ( n_{\textit{cores}} \cdot s_{\textit{die}} + oh_p ) \cdot ec_p$ including the packaging overhead $oh_p$.
\setlength{\abovedisplayskip}{3pt}
\setlength{\belowdisplayskip}{3pt}
\begin{eqnarray}
        ec_p & = & (en_p\cdot ci_{fab} + ec_m + ec_g)/\textit{yield} \\
        ec_{CPU} &= & n_{\textit{units}} \cdot ( ( n_{\textit{cores}} \cdot s_{\textit{die}} + oh_p ) \cdot ec_p + ec_{\textit{base}} )\nonumber
\label{eqEmbCarbonCPU}
\end{eqnarray}
Similar equations are used for RAM, SSD and GPU. The model further includes the embodied carbon of the power supplies, motherboard and enclosure and emissions from assembly.  The model also takes into account the embodied emissions of the infrastructure and of the networking infrastructure, based on estimates worked out in \cite{schneider2023}: for a 1MW data centre, infrastructure embodied carbon is estimated at 1,829 tCO$_2$e on construction, increasing to 2,400 tCO$_2$e after 20 years due to maintenance and replacement of parts of the infrastructure; networking equipment embodied carbon is estimated at 6\% of the total server embodied carbon. We relate this through the individual nodes by scaling with the power consumption of the node, corrected by the data centre PUE. For full details, we refer to the source code and model documentation \cite{hpc_lca_code_wv2025}.

\subsection{Embodied carbon trends}\label{sec:Embodied-carbon-trends}

The main addition to the model for this paper is the incorporation of trends for the evolution of the constituents of the embodied carbon as presented above. 

Consecutive semiconductor process nodes follow an exponential trend,
reflected by Moore's law for the circuit density \cite{schaller2002moore}
and Koomey's law for the energy efficiency \cite{koomey2011web}.
The semiconductor industry aims to keep alignment with those laws,
and therefore the trends in the contributing factors to the embodied
emissions all follow a power laws. For our model, we need trends for the following parameters:

\begin{itemize}[noitemsep,topsep=5pt]
	\item Server thermal design power (TDP) 
	\item Process node embodied carbon per unit area, which gives us trends for CPU and GPU
	\item Solid State Drive (SSD) embodied carbon trend also requires
		\begin{itemize}
			\item SSD number of layers trend
			\item SSD density trend
			\item SSD capacity trend
		\end{itemize}
	\item DRAM embodied carbon trend also requires
		\begin{itemize}[noitemsep,topsep=0pt]
			\item DRAM capacity trend
			\item DRAM density trend
		\end{itemize}
\end{itemize}

For each of these parameters, we perform a non-linear Levenberg-Marquardt least-squares fit \cite{levenberg1944method,marquardt1963analgo}. We use a 95\% confidence interval ($2\sigma$). The fit is judged
on the basis of the sum of the squared differences or 'residuals'
between the input data points and the function values, evaluated at
the same places (SSR, a.k.a. $\chi^{2}$). 

As we are focusing mainly on generative AI data centres, we use the
NVIDIA DGX range of servers, which are widely used both for inference
and training. We use the vendor data sheets from 2013 to 2025 complemented with other databases and sources (listing all source would make the paper too long).

Our ultimate aim is to obtain estimates for the CO$_{2}$ emissions of data centres. These emissions are dominated by integrated circuits. The increase in capacity of a data centre is expressed in terms of power. For every consecutive semiconductor process node, not only the embodied carbon but also the power consumption per unit area changes. We therefore normalise the embodied carbon to the power consumption. We also derive separate trends for embodied carbon of SSDs and DRAM, as their manufacturing processes are different from ICs.

Using the trends for the above parameters, the actual embodied carbon estimate over
time for every constituent $i$ of the server is given by 
\setlength{\abovedisplayskip}{3pt}
\setlength{\belowdisplayskip}{3pt}
\begin{equation}
trend_{EC,i}(\overline{a}_{EC,i},t)=(EC_{i}/EC_{tot})\cdot\textrm{compound}(\overline{a}_{EC,i}-1,t)
\end{equation}
where
\setlength{\abovedisplayskip}{3pt}
\setlength{\belowdisplayskip}{3pt}
\begin{equation}
\textrm{compound}(p,t)=(1+\lvert p\rvert)^{\frac{p}{\lvert p\rvert}t}
\end{equation}
is the generalised compound growth equation. The power consumption trend is likewise 
\setlength{\abovedisplayskip}{3pt}
\setlength{\belowdisplayskip}{3pt}
\begin{equation}
trend_{TDP}(\overline{a}_{TDP},t)=\textrm{compound}(\overline{a}_{TDP}-1,t)
\end{equation}

The final expression for the server embodied carbon trend becomes:
\setlength{\abovedisplayskip}{3pt}
\setlength{\belowdisplayskip}{3pt}
\begin{equation}
trend_{EC,DGX}(t)=\frac{\sum_{i=\textrm{\{SSD,DRAM,CPU,GPU,rest\}}}trend_{EC}(\overline{a}_{EC,i},t)}{trend_{TDP}(\overline{a}_{TDP},t)}\label{eq:trend_dgx}
\end{equation}

Although it is possible to calculate the variance of simple linear expressions using the well-known laws of uncertainty propagation, this is not valid for a complex time dependent model such as Eq.~\ref{eq:trend_dgx} because the exponential growth curve skews the distributions. Therefore, we apply a sampling technique to obtain the 5th, 50th and 95th percentiles. We construct a normal distribution for each of the constituents and sample the compound expression. 

\section{AI data centre growth scenarios}

\subsection{International Energy Agency Scenarios}

\begin{figure}[h]
\includegraphics[width=0.9\columnwidth]{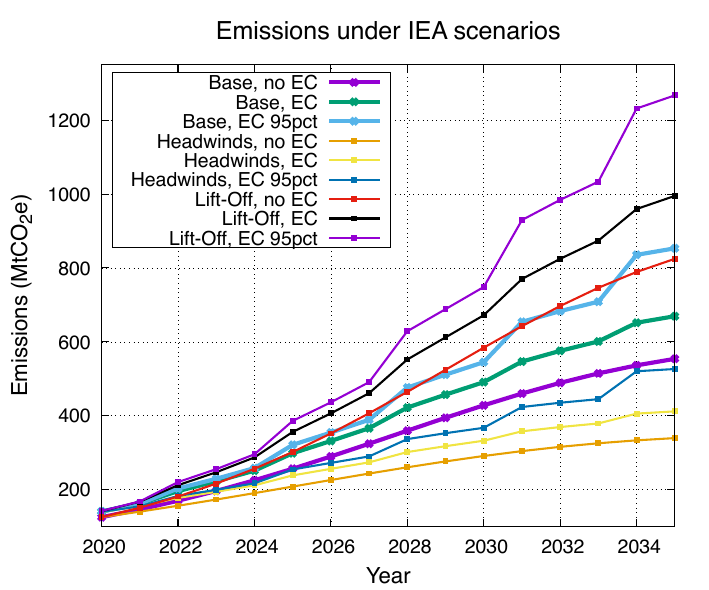}

\caption{Growth in emissions for global AI data centre expansion under IEA scenarios, 3-year server lifetime}\label{fig:emissions-IEA}
\end{figure}

The International Energy Agency (IEA) has published a report with four scenarios for the growth in global data centre electricity consumption (in TWh/y) between 2025 and 2035, using historical data from 2020 to 2025 \cite{iea2025energy}. These scenarios assume logistic growth. The ensuing CO$_2$ emissions calculated using our model are shown in Fig. \ref{fig:emissions-IEA}. We assume a server lifetime of 3 years. This is conservative as GPU servers are replaced within 1.5 to 2 years.

The main observations are that (1) the ``Lift-Off'' scenario would cause and additional 0.8 GtCO$_2$e/y in operational emissions by 2035; (2) the embodied carbon contribution is significant even for median values, 25\% or 0.2 GtCO$_2$e/y; (3) With the 95th percentile estimate for the embodied carbon contribution, the total emissions would be 1.3 GtCO$_2$e/y, i.e embodied emissions make up 40\% of the total. The high spread on the embodied carbon estimate is a result of the exponential growth trends for the constituents.

\subsection{McKinsey Scenarios}

\begin{figure*}[t]
\includegraphics[width=1.5\columnwidth]{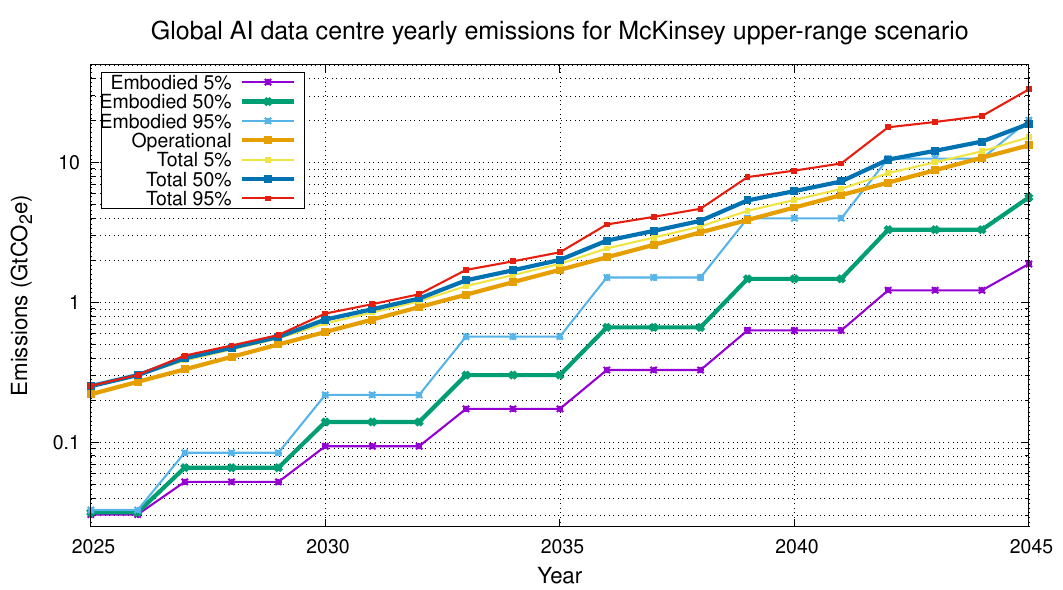}
\caption{Growth in global emissions at 25.9\% CAGR in AI data centre expansion, 3-year server lifetime}\label{fig:emissions-McKinsey-26pct}
\end{figure*}

The CEO of Dell has said that AI would ``drive data centre demand up by 100$\times$ over the next 10 years'' \cite{ref18}. The CEO of OpenAI has said that the world needs 100$\times$ more semiconductor production capacity \cite{ref19}, which amounts to the same.
To show how bad things could get if the industry projections would materialise, we use the McKinsey ``Upper-range'' scenario which project 27\% compound average growth rate (CAGR) in global data centre capacity between 2023 and 2030 \cite{ref3}. We assume a utilisation of 80\% and server lifetime of 3 years. We use 25.9\% CAGR rather than 27\% because this translates to 10$\times$ growth in capacity in ten years and 100$\times$ in 20 years. Fig. \ref{fig:emissions-McKinsey-26pct} shows the resulting emissions. The worst-case estimate yields emissions of about 3 GtCO$_2$e/y by 2035 and more that 30 GtCO$_2$e/y by 2045. 

If this growth was indeed sustained until 2035, it would be already be quite problematic: the total global CO\textsubscript{2} budget for 2035 is 22 GtCO\textsubscript{2}e/y according to the UNEP Emissions Gap Report 2024 \cite{ref16}. Global emissions from electricity generation were 14 GtCO\textsubscript{2}e in 2023 and projected to rise to 15 GtCO\textsubscript{2}e by 2035 without the growth in AI. The projected increase would consume 14\% to the global emissions budget.
But if this trend would persist for 20 years, then emissions from AI data centres alone would exceed the global emissions budget. As we can see from the figure, the worst-case embodied emissions alone would consume that budget.

\section{Conclusion}

We have presented a detailed model of the operational and embodied emissions of AI data centres and have used it to model emissions for the growth scenarios from the IEA and McKinsey. There are two main observations: (1) The embodied carbon contribution is significant in all cases, making up 20\% of total emissions in the median case and up to 40\% in the worst case. The spread is large because the embodied carbon trends follow a power law so small deviations have large effects. (2) Even the emissions under conservative, bounded growth IAE scenarios are of the order of 1 GtCO$_2$e/y and unacceptable in terms of the global CO$_2$ budget. Under the unlimited growth scenarios from McKinsey, which reflect the ambitions of the AI industry, emissions from AI data centres would be downright catastrophic, consuming the entire planetary carbon budget.

\bibliographystyle{IEEEtran}
\bibliography{paper-AI-datacentres}

\end{document}